\documentclass[12pt]{article}       

\usepackage{lipsum}
\usepackage{fancyhdr}
\usepackage{booktabs}
\usepackage{array}
\usepackage{amssymb}
\usepackage{parskip}
\usepackage{hyperref}
\usepackage{enumitem}
\usepackage{mathtools} 
\usepackage{xparse,textcomp}

\usepackage[lofdepth,lotdepth]{subfig}

\usepackage{cite}

\usepackage{parskip}
\usepackage[margin=0.8in]{geometry}

\usepackage{algorithm} 
\usepackage{algpseudocode}

\usepackage{amsmath,amsthm}
\makeatletter

\usepackage[belowskip=-30pt,aboveskip=10pt]{caption} 
\usepackage{amsfonts}
\usepackage{amsopn}

\usepackage{amssymb,amsmath}

\renewcommand\section{\leftskip 0pt\@startsection {section}{1}{\z@}%
	{-3.5ex \@plus -1ex \@minus -.2ex}%
	{2.3ex \@plus.2ex}%
	{\normalfont\Large\bfseries}}

\renewcommand\subsection{\leftskip 0pt\@startsection{subsection}{2}{\z@}%
	{-3.25ex\@plus -1ex \@minus -.2ex}%
	{1.5ex \@plus.2ex}%
	{\normalfont\large\bfseries}}

\renewcommand\subsubsection{\leftskip 0pt\@startsection{subsubsection}{3}{\z@}%
	{-3.25ex\@plus -1ex \@minus -.2ex}%
	{1.5ex \@plus .2ex}%
	{\normalfont\large\bfseries}}
\usepackage{graphicx}

\usepackage{geometry}
\newtheorem{definition}{Definition}

\theoremstyle{remark}

\begin{document}

\title{ \centering{Image Quality Assessment: Learning to Rank Image Distortion Level}}
\author{Shira Faigenbaum-Golovin${^{1*}}$~~Or Shimshi${^2}$ \\
	\small{${^1}$ Duke University, North Carolina, USA}
	\\
	\small{${^*}$ Corresponding author, E-mail address: shira.golovin@math.duke.edu }
	\\
	\small{${^2}$ Tel-Aviv, Israel} 
}

\maketitle
\newcommand{\argmin}{\operatornamewithlimits{argmin}}

\begin{figure}[htbp]
  \centering
  \label{fig:b}\includegraphics[width=\textwidth,height=\textheight,keepaspectratio]{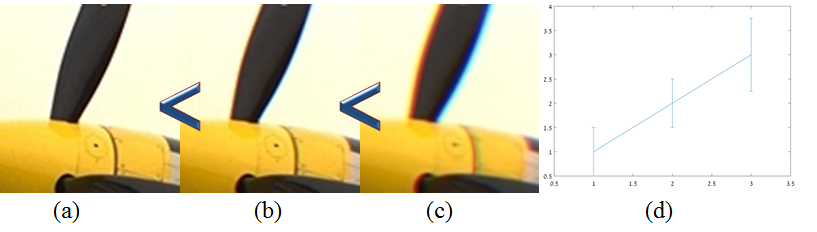}
  \caption{  	Automatically ordered images (a)-(c) from the TID2013 \cite{ponomarenko2015image} by their predicted Lateral Chromatic Aberration level (0, 2, 4 pixels accordingly) using the proposed method in this paper; the “<” sign indicates that the left image is less distorted than the one on the right; (d) a plot of the expected (the x-axis) versus the predicted image order (the y-axis).}

  \label{fig:Fig. 1}
\end{figure}

\begin{abstract}

Over the years, various algorithms were developed, attempting to imitate the Human Visual System (HVS), and evaluate the perceptual image quality. However, for certain image distortions, the functionality of the HVS continues to be an enigma, and echoing its behavior remains a challenge (especially for ill-defined distortions). In this paper, we learn to compare the image quality of two registered images, with respect to a chosen distortion. Our method takes advantage of the fact that at times, simulating image distortion and later evaluating its relative image quality, is easier than assessing its absolute value. Thus, given a pair of images, we look for an optimal dimensional reduction function that will map each image to a numerical score, so that the scores will reflect the image quality relation (i.e., a less distorted image will receive a lower score). We look for an optimal dimensional reduction mapping in the form of a Deep Neural Network which minimizes the violation of image quality order. Subsequently, we extend the method to order a set of images by utilizing the predicted level of the chosen distortion. We demonstrate the validity of our method on Latent Chromatic Aberration and Moire distortions, on synthetic and real datasets.

\end{abstract}

\noindent\textbf{keywords:} Image Quality Assessment, Deep Neural Network,  Image Processing, Dimensional Reduction, Resnet.

\section{Introduction}

Evaluating the quality of an image is a vital task in the domain of image processing. This is crucial for measuring the performance of image processing algorithms, which may unintentionally damage image quality (e.g., a denoising algorithm can reduce the sharpness of the edges). While talking about image quality, often questions arise regarding an image’s quality assessment: like \emph{What is a good image?} And also \emph{Why is image quality assessment so difficult?}  \cite{wang2002image} While the Human Visual System often provides an answer to these questions, it is not efficient and at times prone to subjective judgment. 

In order to avoid repeated evaluation by image quality experts, an automatic image quality procedure is essential. Image quality (IQ) assessment can be addressed either by absolute or relative measure. In the first method, a number representing the IQ of a single image is computed. In the second method, given two images, we indicate which image looks better. It should be noted that usually providing a relative measure is easier, as opposed to evaluating the quality of a single image and scoring it based on its defects. The reason for this is that there is no need to specify what image characteristics influenced the scoring.

Along the years there were various algorithms suggested to crack the enigma of Human Visual System (HVS), which constantly aids humans in this task  \cite{granrath1981role, kleinmann2007investigation}. Those studies paved the way towards designing an objective procedure for image quality evaluation  \cite{faigenbaum2013evaluating, barney2015quantifying, kleinmann2007investigation, zhang2018opinion, van1996working}. However, measuring the perceptual image quality still remains a challenge. In recent years, the interest in image quality metrics was renewed, with the rise of Deep Neural Networks (DNN). The main idea of DNN is that an image quality metric can be defined and evaluated on images. Subsequently, a DNN is constructed to associate between the image and its calculated quality. Recently published papers  \cite{gao2015learning, gao2017deepsim, hou2015blind, xu2016pairwise,talebi2018nima, chetouani2020image, zhu2020metaiqa, ding2020image} demonstrate the benefits of using DNN for IQ evaluation. These methods require the defining and evaluating an absolute image quality metric. This requires the designer to develop a metric and evaluate the distortion as a pre-processing step of DNN. However, the main challenge is that certain image distortions are ill-defined (though they are easy to acquire or simulate), and therefore, no metric exists to evaluate them.
In this paper, we address the question of image quality assessment by introducing a framework for learning \emph{relative IQ}. We look on image distortion as the degradation of the ideal image or as a deviation from the “perfect” image. Thus, we propose a relative-order-preserving image quality, and bypasses the challenge of defining the desired distortion. 

Our method was inspired by  \cite{vendrov2015order}, in which the semantic hierarchy of words, and sentences was learned. Although this particular hierarchy is based on the hypernymies of words, for our purposes, the aim is to maintain the order of images based on their quality. In what follows, we introduce our relative IQ method (subsection \ref{sec:IQPDR}) and later it is extended to rank a set of images (subsection \ref{sec:IQR}). Then, in section \ref{sec:IQOPF}, we describe the construction of the training, validation, and testing datasets. The validity of our method is demonstrated through the order-preserving dimension reduction of two distortions: Chromatic Aberration and Moire (section \ref{sec:IQ_results}). The paper concludes with a discussion of future directions for methodological enhancements (section \ref{sec:summary_and_future}).

\section{Proposed Method}
\subsection {Ordering Image Pair by Distortion Level}
\label{sec:IQPDR}
First, we introduce the relative image quality measure in reference to the question\textit{ “given two images, \textup{$A$}, and  \textup{$B$} as well as image distortion \textup{$d$}, is image  \textup{$B$} more distorted than image \textup{$A$} with respect to \textup{$d$}”?} We answer this question by defining image quality order with respect to the selected distortion. Subsequently, we look for a dimension reduction mapping from the image domain to a the natural numbers, which maintain the IQ order. This mapping is later utilized to make order in an unseen image pair. In mathematical terms, we define the \textit{IQ-order of two given Regions of Interest} (ROI) of an image as:

{\setlength{\parindent}{0cm}
\begin{definition}\label{def:def1}
Let $R_A$ and $R_B$ be two registered ROIs, and let $d$ be the distortion we would like to evaluate. Then, \textbf{ROI-IQ-order} is defined as $R_A$ $<_d$ $R_B$, meaning that ROI $R_A$ is less distorted then ROI $R_B$, with respect to $d$. 
\end{definition}
Let us now extend this definition to the entire image:
\begin{definition}\label{def:def2} 
Let $A$ and $B$ be two images, as well as the set of their registered ROIs $\{R_{A,i}\}$, $\{R_{B,i}\}$.  If  the following condition stands \#\{ i $|$  $R_{A,i} <_d R_{B,i}$\} $>$ \#\{ i $|$  $R_{A,i} >_d R_{B,i}$\} then we define \textbf{IQ-order} as $A <_d B$. I.e. image $A$ is less distorted then image $B$ with respect to the tested ROIs, and the selected distortion.
\end{definition}

Next, we define IQ-order-preserving mapping as  
\begin{definition}
\label{def:def3}
Let $A$ and $B$ be two images, such that $A <_d B$ and let $S$=\{$\langle R_{A,i}, R_{B,i} \rangle $ $|$ $R_{A,i} <_d R_{B,i}$ \}$_{i=1..N}$ be their set of ordered registered ROI pairs, and $f \colon \mathbb{R}^{n \times m} \rightarrow \mathbb{R}$ a dimensional reduction function. We say that the mapping $f$ is \textbf{IQ-order-preserving} if for any ROI pair in $S$, $f$ is order-preserving. That is if $\forall i$ $R_{A,i} <_d  R_{B,i}  \Longrightarrow  f( R_{A,i}) < f(R_{B,i})$.
\end{definition}
}

\noindent Now, image quality question can be formulated with respect to the order-preserving mapping.

\noindent \textbf{Problem definition:} \hspace{2mm} Let $A, B$ be two images such that $A <_d B$ and also let $S$=\{$\langle R_{A,i}, R_{B,i} \rangle $ $|$ $R_{A,i} <_d R_{B,i}$ \}$_{i=1..N}$ be set of ordered registered ROI pair set. The IQ-order-preserving mapping $f$, is found such that it will minimize the image quality order violation
\begin{equation} \label{eq:1}
f =\operatorname*{argmin}_{f \colon \mathbb{R}^{n \times m}  \rightarrow\mathbb{R}} \frac1{N}  \sum_{(R_{A,i},R_{B,i} )\in S} E(R_{A,i},R_{B,i} )
\end{equation}
where the loss function, $E$, for an ordered pair $\langle R_A, R_B \rangle $ is defined as 
\begin{equation}
E(R_A,R_B)   = \textup{max}(0, (f(R_A) + \epsilon) - f(R_B))^2
\end{equation}

Once the optimal $f$ is found (see subsection \ref{Network_Architecture} for details), given a pair of images we can find the less distorted image using definition \ref{def:def3}.

\subsection{Ranking the Distortion of Image Set} 
\label{sec:IQR}
The definition of ''order'' on image pair can be extended to rank image set, utilizing their distortion level. Once the mapping $f$, which minimizes equation \eqref{eq:1}, is found, it can be used to calculate the relative score of images. As a result, one can order a given image set with respect to the values of $f$. Specifically, for a set of images $ \{A_j\}_{j=1..J}$, with the corresponding ROIs $R_{j,i}$, we calculate $f(R_{j,i})$. Subsequently, the ranking of this set with respect to a given distortion is achieved by ordering the values $f(R_{j,i})$ for each specific ROI index ($i$), and then calculating the median af the ranking across all the image patches ($j$).

We illustrate this procedure in the following example. We rank four images using three patches with given predicted IQ values. Table \ref{tab:table_example} illustrates all the steps of the process (a) starting with predicting the IQ values using some learned $f$ (each row in the matrix appearing in the first column correspond to a different patch, and each column to different image), (b) ranking their ROI’s and later (c) ranking the four images. We demonstrate this concept in real case scenario by ranking images with respect to Chromatic Aberration distortion in Figure \ref{fig:Fig. 1}.  Images (a)-(c) are ordered according to their rank, and (d) is a plot of the expected ranking versus the predicted one.

\begin{table}[H]
\centering
\caption{Example of ranking four image, with respect to their three patches}
\label{tab:table_example}
\begin{tabular}{lllll}
\cline{1-3}
\multicolumn{1}{c}{Predicted IQ values}                                                                      & \multicolumn{1}{c}{Calculated patches Ranks}                                                               & \multicolumn{1}{c}{Calculated images rank}   &  &  \\ \cline{1-3}
\multicolumn{1}{c}{\begin{tabular}[c]{@{}c@{}}{[}1 2 3 4;\\ 4 8 9 12;\\ 2 3 5 4{]}\end{tabular}} & \multicolumn{1}{c}{\begin{tabular}[c]{@{}c@{}}{[}1 2 3 4;\\ 1 2 3 4;\\ 1 2 4 3{]}\end{tabular}} & \multicolumn{1}{c} {{[}1 2 3 4{]}} &  &  \\ \cline{1-3}
\end{tabular}
\end{table}
\vspace{-12mm}

\subsection {Network Architecture}
\label{Network_Architecture}
In this study we design the order-preserving mapping as a Deep Neural Network (DNN). The network architecture comprises a Siamese network of a pair of ResNet architecture \cite{he2016deep}, each performing a dimension reduction. Later, we calculate the loss function which maximizes the distance between mismatches of the dimension reduction via equation \ref{eq:1} (which is also called squared negative smoothed hinge loss (SNSHL) \cite{rennie2005loss}. See table \ref{tab:table_arch} for detailed  network architecture.

\begin{table}[H]
\centering
\caption{Architecture of the IQ order-preserving network. Building blocks are shown in brackets (and consists of three consequent ReLU’s), with the numbers of blocks stacked. The network input is two concatenated color patches of $32 \times 32$.}
\label{tab:table_arch}

\begin{tabular}{|l|l|l|l|}\hline

\textbf{Layer name}                                                             & \textbf{Output size} & \multicolumn{2}{l|}{\textbf{Order-preserving Net104-layer}}                                                     \\ \hline
\textbf{slice}                                                                  &                      & \multicolumn{2}{l|}{slice point 3}                                                                                                                     \\ \hline
conv1                                  & $16 \times 16 \times 2$            & $7 \times 7, 64$, stride 2                                                   & $7 \times 7, 64$, stride 2                                                  \\ \hline
conv2.x                              & $8 \times 8 \times 2$                & $\left[ \begin{array}{cc}1 \times 1, 64 \\3 \times 3, 64 \\1 \times 1, 256 \end{array} \right] \times 3$   & $\left[ \begin{array}{cc}1 \times 1, 64 \\3 \times 3, 64 \\1 \times 1, 256 \end{array} \right] \times 3$                  \\ \hline
conv3.x                              & $4 \times 4 \times 2 $               & $\left[ \begin{array}{cc}1 \times 1, 128 \\3 \times 3,128 \\1 \times 1, 512 \end{array} \right] \times 4$   & $\left[ \begin{array}{cc}1 \times 1, 128 \\3 \times 3,128 \\1 \times 1, 512 \end{array} \right] \times 4$                  \\ \hline
conv4.x                              & $2 \times 2 \times 2$                & $\left[ \begin{array}{cc}1 \times 1, 256 \\3 \times 3, 256 \\1 \times 1, 1024 \end{array} \right] \times 6$   & $\left[ \begin{array}{cc}1 \times 1, 256 \\3 \times 3, 256 \\1 \times 1, 1024 \end{array} \right] \times 6$                   \\ \hline    
conv5.x                              & $1 \times 1 \times 2 $               & $\left[ \begin{array}{cc}1 \times 1, 512 \\3 \times 3, 512 \\1 \times 1, 2048 \end{array} \right] \times 3$   &   $\left[ \begin{array}{cc}1 \times 1, 512 \\3 \times 3, 512 \\1 \times 1, 2048 \end{array} \right] \times 3$                \\ \hline
conv1                                 & $1 \times 1 \times 2 $ 		& $1 \times 1, 1$, stride 2   &  $1 \times 1, 1$, stride 2               \\ \hline
\begin{tabular}[c]{@{}l@{}}squared negative \\ smoothed hinge loss\end{tabular} & $1 \times 1 \times 1$                &  $1 \times 1, 1$, stride 2  &   $1 \times 1, 1$, stride 2          \\ \hline                 
\end{tabular}
\end{table}

\subsection{Accuracy Evaluation of the Predicted Order}
\label{accuracy}
Measuring predicted order accuracy can be separated into two scenarios (a) for a pair of images, (b) for image set. While the first one can be evaluating as the True Positive (TP) percentage. The accuracy of the latter one, is performed using the Spearman's correlation coefficient \cite{lehmann1975nonparametrics}, which is widely utilized to detect trends in data - given reference data. Thus, given two measurements vectors $x$, $y$, and their corresponding ranks, $r_x$ and $r_y$ the Spearman's correlation coefficient is calculated as the  
$$\rho=\frac {cov(r_x,r_y)}{\sigma_{r_x}\sigma_{r_y }},$$ 
where $\sigma_{r_x}, \sigma_{r_y}$ are the standard deviations of the rank variables.

Subsequently, for a patch $i$, we apply the Spearman's correlation of the ranks of the predicted values $f(R_{j,i})$,  and a monotonically increasing sequence with equal length (result in a correlation coefficient $\rho_i$). The accuracy of the predicted image set ranking is the median of $\rho_i$ across all image patched. In the example above, the correlation coefficients of the patches ranking are [1 1 0.8], with $median(\rho_i)=1$. Therefore, we conclude that there is a monotonicity trend in the data, and the rank prediction is perfect.

\section{Database Creation}
\label{sec:IQOPF}
\begin{figure}[htbp]
  \centering
  \label{fig:c}\includegraphics[width=\textwidth,height=\textheight,keepaspectratio]{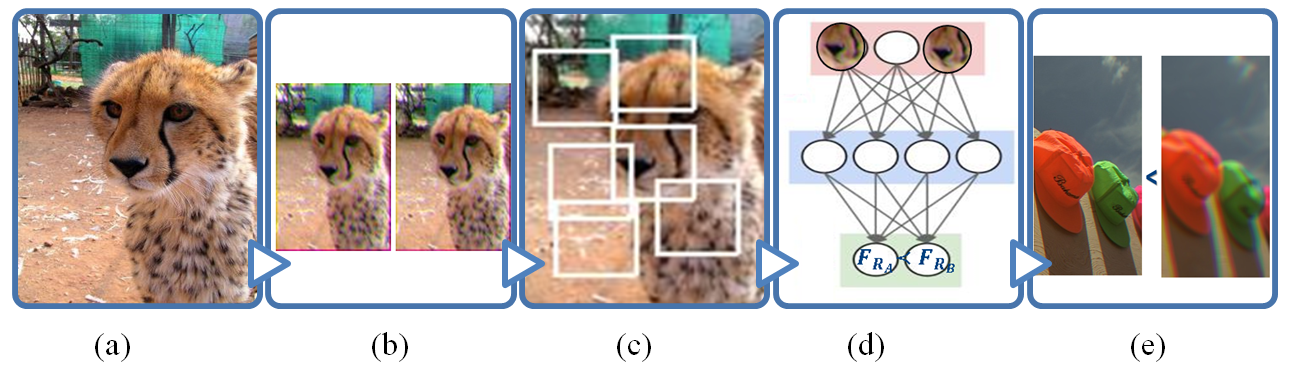}
  \caption{Our experimental flow: (a) Acquire a database of images with a chosen distortion (subsection \ref{sec:Database}) (b) create pairs of images each corresponding to different levels of the chosen distortion (subsection \ref{sec:Distortion})  (c) find areas which correspond to high values of the distortion, and cropping ROIs of size 32x32 (subsection \ref{sec:Extraction})  (d) learn the order-preserving dimension reduction and (e) test it on new datasets.}
  \label{fig:Fig. 2}
\end{figure}

Our experimental flow consisted of the following steps: (a) acquiring a dataset of images for training and validation; (b) pairing or simulating images of the same scene, each corresponding to different levels of distortion, and accompanied with a predefined IQ-order; (c) extracting ROIs which contain the desired distortion; (d) learning the order-preserving dimensional reduction function, f; and (e) executing the method on various datasets. The general flow is illustrated in Figure \ref{fig:Fig. 2}.

\subsection{Database Acquisition}
\label{sec:Database}

We tested our methodology on learning to rank the Lateral Chromatic Aberration and Moire distortions. Since no dataset of pair of images, accompanied by the level of distortion was available for training-testing purposes, we created our own dataset by simulating a distortion on existing images (for the Lateral Chromatic Aberration case) or by creating synthetic images (for the Moire case). For the first case we took an existing set of images (ImageNet dataset \cite{deng2009imagenet}) as a baseline and for each image in the dataset we created a pair of images, each with a random level of distortion. For the Moire case, we first created images with repetitive pattern, and then created an image pair with different distortion levels. More details on how each distortion was simulated can be found below. After the model was trained, we tested it on a set of real images of TE42.v2 chart (designed and produced by Image Engineering \cite{TE42}).

\subsection{Distortion Simulation}
\label{sec:Distortion}

\begin{figure}[htbp]
  \centering
  \label{fig:e}\includegraphics[width=\textwidth,height=4.5cm,keepaspectratio]{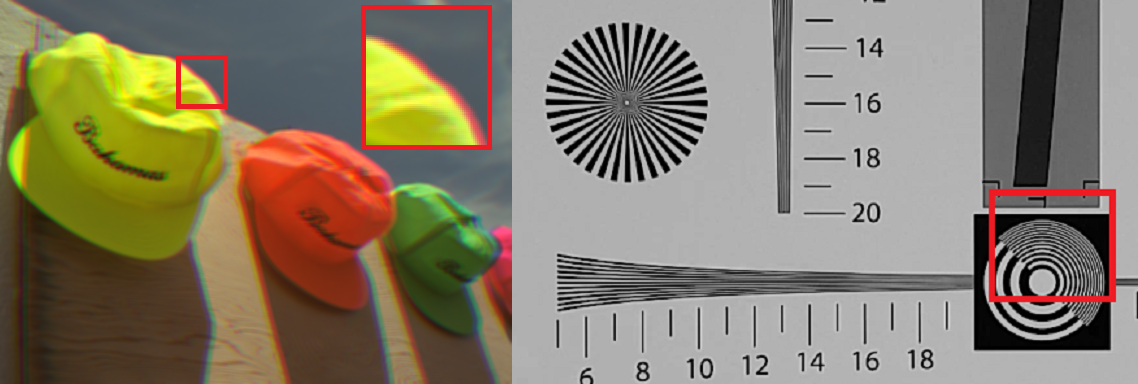}
  \caption{Left: Image with Chromatic Aberration distortion form the TID2013 dataset [15]; right: Resolution chart with Moire effect. Marked in red are the areas with the desired distortion.}
  \label{fig:Fig. 4}
\end{figure}

We verified the validity of the proposed order-preserving method on \textit{Lateral Chromatic Aberration} (LCA) and \textit{Moire} distortions (Figure \ref{fig:Fig. 4}). The \textbf{LCA} distortion appears when the colors convergence point is not unique (which stems from a failure of a lens to focus). This effect is especially seen as a blur and “rainbow” edge in areas of contrast. The LCA dataset, utilized by our method, was constructed by distorting the ImageNet dataset \cite{deng2009imagenet}, where the RGB channels of each image of the dataset  were shifted with a random shift of size  $\sim U(1, 5)$ pixel, in one of the square diagonal directions. The \textbf{Moire} distortion (or aliasing) is an effect that causes different signals to become indistinguishable when sampled \cite{sidorov2002suppression}. It occurs upon the existence of repetitive patterns of high spatial frequencies, which are sampled with different frequency. Since natural images usually do not depict constant frequency which could serve for training Moire distortion, we had to create synthetic image dataset with constant high-frequency patterns. Our dataset contained the following simulated repetitive patterns (i.e. resolution bars, resolution net, Siemens-star, resolution wedges, concentric rings). In order to simulate the Moire effect we used image resize with bicubic interpolation without antialiasing option, with randomly sampled resize factor  $\sim U (1.5, 10)$.

\subsection{ROIs Extraction}
\label{sec:Extraction}
Once IQ-Order-preserving image pair $\langle A,B\rangle$ is created (by applying two random levels of the chosen distortion) the IQ-Order-preserving ROIs of size 32x32 are extracted. The ROIs pairs $\langle R_{A,i}, R_{B,i} \rangle$, are chosen as the ones corresponding to the maximal values of the error map: $ErrMap= |A-B|$. We choose patches with a sufficient amount of distortion (in our experiments we choose ROI pair with $ErrMap$ larger the $0.025\%$ of the image area. 

As a result, our constructed dataset for each distortion (LCA and Moire) consisted of about 6 Million image patches.

\section{Experimental Results}
\label{sec:IQ_results}

We demonstrate the validity of the proposed method by learning two order preserving mapping, each corresponding to different distortion (either the LCA and Moire distortions). We trained a DNN, described in subsection \ref{Network_Architecture}, using a dataset of ordered image pairs (discussed in section \ref{sec:IQOPF}). The optimal ordering mapping was later used to (a) predict the IQ order of image pairs, and later (b) to ranked registered sets of images on new datasets.

While the training and testing of the DNN performed well on synthetic data (Table \ref{tab:sum}), the remained question was \textit{what is the accuracy of using the model, which was trained on synthetic data, for real life images?}. Unfortunately, real life dataset of distortions, with images rank was not available. Therefore, we acquire a new dataset specifically for this task, which consisted of two tests sets of images of TE42 version2 chart \cite{TE42}. One set depicted the TE42 chart in various LCA levels, and other contained images with different Moire effect (about 13 images in each set for each distortion). Subsequently, the images were ranked by an independent image quality expert using his human vision system. Later, using we sampled different patches from the images using Monte Carlo to enrich our dataset. In order to simulate different rank test, each pair was randomly cropped to a size of 150x150 pixels, while depicting the desired distortion. For each cropped pair, we predicted the IQ metric and ranked the cropped patches. The results are summarized bellow, and in Table \ref{tab:sum}. 

In addition, we also we used another dataset, entitled TID2013 dataset \cite{ponomarenko2015image}), to evaluate the performance of image set ranking. This dataset consist of varios images, each undergo under a set of image distortion, in varios levels.

\subsection{Accuracy of Ranking Image Pair }
Our trained models resulting in 97\%, 94\% TP percentile for the simulated test data of the LCA and Moire distortions accordingly. Later, we used the images of the TE42 chart for evaluating the order prediction in real life scenario. Having the 13 images, We simulated 150 different Monte Carlo experiments - by sampling \textbf{pairs} of images from the test set, which contain the required distortion. For each cropped pair, we predicted the IQ metric and ranked the cropped patches. This experiment  resulted in TP percentile of 80\% and 85\% for LCA and Moire for a real life images. This result indicate that although the model was trained on synthetic data, it performs well on real life images as well.

\subsection{Accuracy of Ranking Image Set}
\begin{figure}[H]
  \centering
  \label{fig:f}\includegraphics[width=\textwidth,height=5.6cm,keepaspectratio]{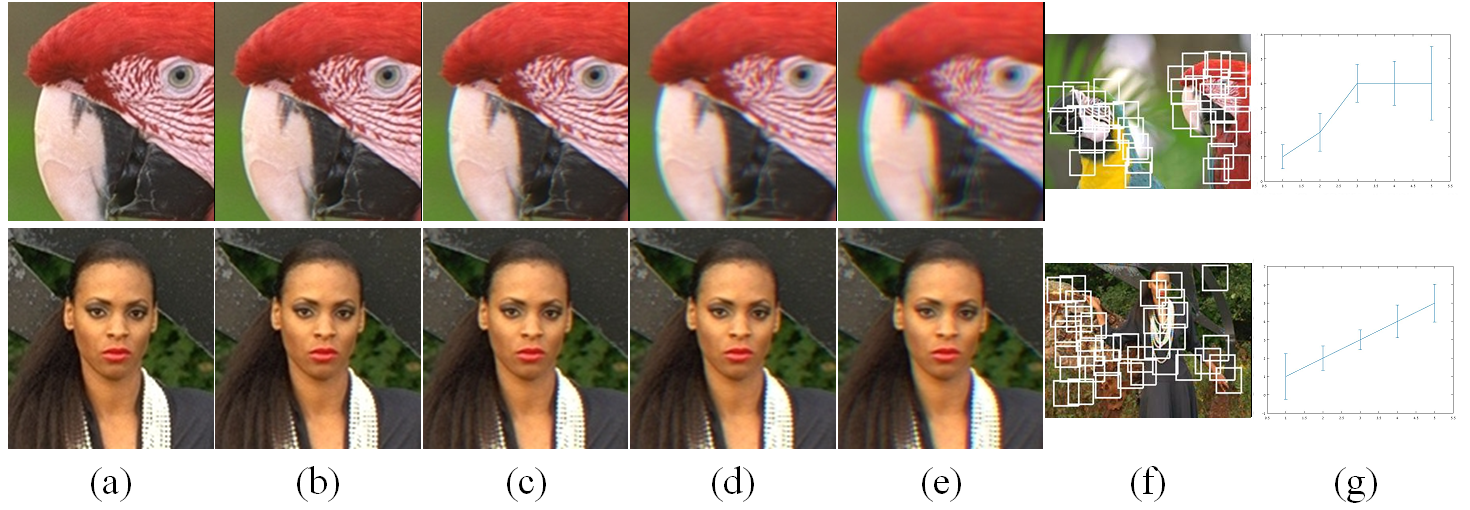}
  \caption{Zoomed in Images ordered by the rank of the predicted IQ measure from the TID2013 dataset (a)-(e). Inspected ROIs, chosen via error map described in subsection \ref{sec:Extraction} (f), a graph with the expect (the x-axis) versus predicted rank values (the y-axis) (g). The accuracy correlation of those sets is $\rho=0.8,1$ (from top to bottom row).}
  \label{fig:Fig. 5}
\end{figure}

Later, we turned to ranking set of image using the methodology described in \ref{sec:IQR}. Our first test was performed on the TID2013 dataset\cite{ponomarenko2015image}, using images created with chromatic aberrations (distortion marked as No. 23 in the dataset). The dataset contained images with chromatic aberration levels 1-5, unfortunately, it is not specified what levels of LCA were used for TID2013 creation. Based on our examination Level 5 of LCA in TID2013 was more than 5 pixels (the value used in our training set), therefore we didn't use it in the ranking procedure. It should be noted that the ROIs were selected by calculating the error map as described above. Since, the images were labeled according to the distortion level, we could check our prediction accuracy against the expected one. The median ranking for the 25 image sets resulted in accuracy correlation of $\rho=1$. Examples of order-preserving ranking can be seen in Figure \ref{fig:Fig. 1} and Figure \ref{fig:Fig. 5}. We see that the distortion level of the images with the airplane (\ref{fig:Fig. 1}) and the one with the lady was predicted perfectly, while for the case of the parrot the scoring was almost good (with images(c)-(e) received the same score).

\begin{figure}[H]
  \centering
  \label{fig:g}\includegraphics[width=\textwidth,height=\textheight,keepaspectratio]{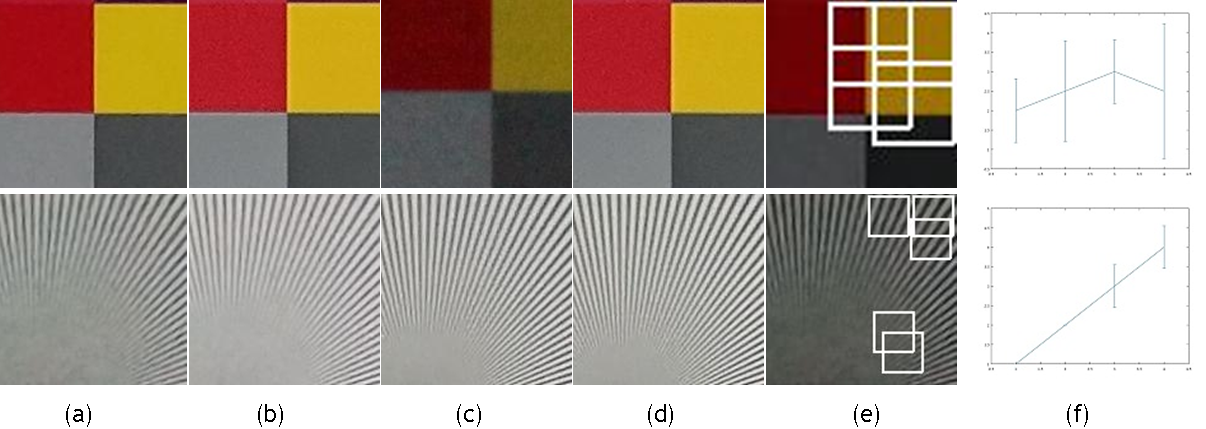}
  \caption{Example of quadruplets, ordered by the rank (a-d), ROIs used for ranking (e), a graph with the expect (the x-axis) versus the median predicted rank values (the y-axis) (f). In the first row, LCA distortion was measured - resulted with accuracy of 0.9. In the second row Moire was assessed, and resulted with accuracy of 1.}
  \label{fig:Fig. 6}
\end{figure}

Subsequently, we tested our ranking methodology on the ordered images set depicting the TE42.v2 chart (discussed above). We simulated 150 image sets by we sampling quadruplets of images from the original dataset and cropped them randomly to a size of 150x150 pixels while depicting the desired distortion. For each cropped quadruplet, we predicted the IQ metric and ranked the cropped patches. The experiment resulted in a median accuracy of $\rho=0.7$ and $\rho=0.8$ for LCA and Moire. We provide two examples of the order prediction for the two distortions in Figure \ref{fig:Fig. 6}). We see that for the LCA predicted order was almost perfect, while for the Moire the order was flawless. This example demonstrate that on real life images the proposed methodology works well as well.

\begin{table}[H]
\centering
\caption{Summary of order-preserving image quality assessment tests for LCA and Moire distortions added artificially or newly acquired in real images}
\label{tab:sum}
\begin{tabular}{lll}
\hline
\textbf{Test}                    & \textbf{LCA} & \textbf{Moire} \\ \hline
Synthetic data, test set (pairwise ordering \%TP)  & 97\%         & 94\%           \\
TE42 chart (pairwise ordering \%TP)                & 80\%         & 85\%           \\
TID2013 (image set ordering $\rho$) & 1            & N/A            \\
TE42 chart (image set ordering $\rho$)              & 0.7          & 0.8            \\ \hline
\end{tabular}
\end{table}

\section{Summary and Future Directions}
\label{sec:summary_and_future}
Assessing image quality is a key problem when evaluating image processing algorithms. The challenge can be addressed either by absolute or relative measurements. As the absolute metric is sometimes ill-defined, and can be harder to implement, a relative metric can solve the problem. In this paper we suggested a relative method, which looks for an optimal mapping that maintains the order of pair of images. Namely, given a pair of images, the mapping returns a pair of scalars that are ordered based on IQ. Subsequently, we proposed extending the mechanism for ranking a set of registered images. The ranking was performed by ordering the images by the found function values.

We demonstrated the validity of our method by constructing a Deep Neural Network and testing it on two distortions: Chromatic Aberration and Moire. The test’s accuracy on synthetic data as well as real data showed satisfactory results. Our test demonstrates that even though the training was performed on synthetic data, the results achieved on real data was satisfactory. In addition, initial experiments showed the potential of utilizing the infrastructure for edge roughness and sharpness assessment. Our method paves the way towards learning to measure a wide range of image distortions.


\section{Acknowledgments}
We would like to thank Dr. Shay Maymon and Dmitry Paus for their insightful and valuable comments and suggestions. We would also like to thank Dmitry Grilikhes and Evgeny Bespechansky. Shira is grateful to the Eric and Wendy Schmidt Fund for Strategic Innovation, and to the Zuckerman-CHE STEM Program for supporting her research.

\bibliographystyle{spmpsci}
\bibliography{references3}
\end{document}